\documentclass[letterpaper, 10 pt, conference]{ieeeconf}  % Comment this line out
\IEEEoverridecommandlockouts                              % This command is only
\usepackage{graphics} % for pdf, bitmapped graphics files
\usepackage{epsfig} % for postscript graphics files
\usepackage{amsmath} % assumes amsmath package installed
\usepackage{amssymb}  % assumes amsmath package installed
\usepackage{booktabs}
\usepackage{units}
\usepackage{accents}
\usepackage{prettyref} %\newrefformat{ch}{\mbox{chapter \ref{#1}}}
\newrefformat{sec}{\mbox{section \ref{#1}}}
\newrefformat{tab}{\mbox{Tab. \ref{#1}}}
\newrefformat{fig}{\mbox{Fig. \ref{#1}}}
\newrefformat{algo}{\mbox{Algorithm \ref{#1}}}
\usepackage{color}
\usepackage[super,negative]{nth}
\usepackage{tikz}
\usetikzlibrary{shapes,arrows}
\usepackage{pgfplots} 
\usepackage{pgfgantt}
\usepackage{pdflscape}
\pgfplotsset{compat=newest} 
\pgfplotsset{plot coordinates/math parser=false}
\newlength\fwidth
\usepackage{grffile}
\usetikzlibrary{plotmarks}
\usetikzlibrary{arrows.meta}
\usepgfplotslibrary{patchplots}
\usepackage[font=scriptsize]{caption}
\usepackage{subcaption}
\usepackage{stmaryrd}

\usepackage{amsthm}
\usepackage{thmtools}
\usepackage{soul}

\newcommand{\alex}[1]{{#1}}

\newcommand{\mathmin}{\operatorname*{minimize}}
\newcommand{\mathst}{\text{s.t.}}
\newcommand{\pplus}{\scriptscriptstyle +}

\newcommand{\T}{\scriptscriptstyle\top}       % Transpose

\declaretheoremstyle[headfont=\bfseries]{normalhead}
\declaretheorem[style=normalhead]{assumption}
\declaretheorem[style=normalhead]{remark}

\title{\LARGE \bf
Nonlinear Model Predictive Control for Distributed Motion Planning in Road Intersections Using PANOC
}

\author{Alexander Katriniok, Pantelis Sopasakis, Mathijs Schuurmans, Panagiotis Patrinos% <-this % stops a space
	\thanks{A. Katriniok is with Ford Research \& Innovation Center, 52072 Aachen, Germany, {\tt\small de.alexander.katriniok@ieee.org}.}
	\thanks{P. Sopasakis is with Queen's University Belfast, School of Electronics, Electrical Engineering and Computer Science, Centre For Intelligent Autonomous Manufacturing Systems (i-AMS), Belfast, Northern Ireland, UK, {\tt\small p.sopasakis@qub.ac.uk}.}
	\thanks{M. Schuurmans and P. Patrinos are with the Department of Electrical Engineering (ESAT-STADIUS), KU Leuven, 3001 Leuven, Belgium, {\tt\small mathijs.schuurmans@kuleuven.be, panos.patrinos@esat.kuleuven.be}.}}%
\begin{document}

\maketitle
\thispagestyle{empty}
\pagestyle{empty}

%%%%%%%%%%%%%%%%%%%%%%%%%%%%%%%%%%%%%%%%%%%%%%%%%%%%%%%%%%%%%%%%%%%%%%%%%%%%%%%%
\begin{abstract}
The coordination of highly automated vehicles (or agents) in road intersections is an inherently nonconvex and challenging problem. In this paper, we propose a distributed motion planning scheme under reasonable vehicle-to-vehicle communication requirements. Each agent solves a nonlinear model predictive control problem in real time and transmits \alex{its} planned trajectory to other agents, which may have conflicting objectives. The problem formulation is augmented with conditional constraints that enable the agents to decide whether to wait at a stopping line, if safe crossing is not possible. The involved nonconvex problems are solved very efficiently using the proximal averaged Newton method for optimal control (PANOC). We demonstrate the efficiency of the proposed approach in a realistic intersection crossing scenario.
\end{abstract}

%---------------------------------------------------------------------------
%---------------------------------------------------------------------------
%---------------------------------------------------------------------------
% MAIN PAPER CONTENT 
%---------------------------------------------------------------------------
%---------------------------------------------------------------------------
%---------------------------------------------------------------------------
%!TEX root = ./main.tex
%---------------------------------------------------------------------------
%---------------------------------------------------------------------------
%---------------------------------------------------------------------------
% INTRODUCTION
%---------------------------------------------------------------------------
%---------------------------------------------------------------------------
%---------------------------------------------------------------------------
\section{Introduction}
\label{sec:introduction}

\alex{Automated vehicles (AV) today either operate in a \textit{reactive} way by solely basing their actions on sensor readings or exploit an uncertain estimate of future motion trajectories of other road users to be more \textit{proactive}. With vehicle-to-vehicle (V2V) communication, more reliable information about future trajectories of the surrounding traffic can be adopted to operate the AV more efficiently \cite{Hobert2015}. Especially in intersections, throughput could significantly be improved when, instead of using traffic lights or signs, vehicles would negotiate intersection crossing using V2V communication.}
	
%Highly automated vehicles (AV) today commonly operate in a pure \textit{reactive} way, i.e., they perceive their environment via sensors and react accordingly, often without planning. With information about the future behavior of the surrounding traffic, these vehicles could be operated even more efficiently as by  anticipating and act in a more \textit{proactive} fashion. Especially in intersections, throughput could significantly be improved when, instead of using traffic lights or signs, vehicles would negotiate intersection crossing. 

In this paper, we address \alex{the} control problem \alex{of coordinating vehicles (referred to as \textit{agents}) in intersections} by means of distributed nonlinear model predictive control (MPC), assuming that V2V communication is available for information exchange. \alex{To this end}, optimization-based strategies such as MPC appear to be appropriate to deal with constrained motion planning \alex{problems} incorporating anticipated trajectories of conflicting agents. 

In recent years, MPC has been widely used for this purpose. 
The authors in \cite{Campos2014} introduce a decentralized control scheme which relies on the solution of two convex quadratic programs (QP) to determine the order in which the agents cross the intersection. An MPC-based scheme is proposed in \cite{Hult2016}, which utilizes a central coordination unit to solve a high-level time slot allocation optimal control problem (OCP), while agent controls are determined as part of a nested low-level OCP. Essentially, every agent solves a QP and two linear programs (LP) and transmits the optimization results to the central coordinator which solves the high-level nonlinear program (NLP). This is extended in \cite{Shi2018} with rear-end collision avoidance. A similar scheme is proposed in \cite{Kneissl2018} where a centralized coordination unit is in charge of time slot allocation, while agents are controlled in a decentralized fashion and all involved OCPs are convex QPs. Moreover, \cite{Molinari2018} presents a decentralized consensus-based control strategy which determines the intersection crossing order as part of a high-level consensus algorithm and solves a convex optimization problem to determine vehicle controls on the lower level. Instead of using time slots, collision avoidance is ensured by bounding distances between agents from below. In \cite{Katriniok2017a}, the authors have outlined a fully distributed MPC scheme where every agent solves a nonconvex quadratically constrained QP (QCQP). Instead of utilizing time slots, agents must keep a minimum distance to each other \alex{--- defined with respect to their joint collision point}.% in the intersection.   

%---------------------------------------------------------------------------
%---------------------------------------------------------------------------
\subsection{Main Contribution and Outline}
\label{sec:introduction_contribution} 

\alex{To orchestrate AVs in intersections}, we rely on a distributed MPC scheme in which every agent solves its nonconvex optimal control problem simultaneously and broadcasts the optimized trajectories to the other agents via V2V communication. We believe that a distributed scheme is more flexible, resilient and scalable than a centralized one when it eventually comes to in-vehicle implementation.

In the literature, most control concepts rely on a formulation in which collision avoidance is tightly related to the intersection scenario, that is, through time slots or collision points \cite{Campos2014,Hult2016,Katriniok2017a}. \alex{For AV motion control, it is desirable to utilize a single controller for multiple scenarios. Therefore, we take a first step to generalize our problem formulation in \cite{Katriniok2017a} to be able to cover a wider range of use cases.}
%Consequently, these concepts might not necessarily be applicable outside intersections. Therefore, we generalize our problem formulation in \cite{Katriniok2017a} to cover a wider range of use cases with a single controller. 
Instead of using joint collision points along the agents' path coordinate, we propose to formulate collision avoidance in a Cartesian frame by mapping the agents' path coordinate to their respective global \alex{Cartesian} coordinates. The authors in \cite{Molinari2018} are pursuing this direction using a linear mapping. By using B-spline functions, we allow for nonlinear mapping functions which are necessary to describe arbitrary realistic driving maneuvers. Eventually, we utilize the area overlap of the agents' bounding boxes (plus some safety margin) to define collision avoidance (CA) constraints. This formulation can easily be applied independent from intersections, e.g., for rear-end collision avoidance or lane change maneuvers. Moreover, we introduce a methodology to embed \textit{conditional constraints} into the motion planning problem, such as waiting at the stopping line if safe intersection crossing is impossible. Finally, we propose to exploit the proximal averaged Newton-type method for optimal control (PANOC) \cite{Sathya2018,Stella2017} to solve the  resulting nonconvex NLP in real-time. 

The remainder of the paper is organized as follows: Section \ref{sec:modeling} introduces the vehicle kinematics model along with the underlying assumptions and the relationship between local and global coordinate frames. The distributed motion planning problem is then formulated in \prettyref{sec:problem}, while \prettyref{sec:numsol} outlines how an efficient numerical solution is obtained. Simulation results are finally discussed in \prettyref{sec:results}. 

\subsection{Notation}
Hereafter, $x_{k+j\mid k}$ will stand for the prediction of variable $x$ at the future time step 
$k+j$ given information up to time $k$. 
For $x\in\mathbb{R}^n$ and $i\in\{1,\ldots, n\}$, $[x]_i$ is the $i$-th entry of $x$.
In addition, $\mathbb{N}^+$ is the set of positive integers, $[x]_{\pplus} \triangleq \max\{x,0\}$
and  \(A^{\T}\) denotes the transpose of a matrix \(A{}\in{}\mathbb{R}^{m\times n}\).

%=========================================================================
%=========================================================================
% MODELING
%=========================================================================
%=========================================================================
\section{Modeling}
\label{sec:modeling}

We rely on the following fundamental assumptions: 
\begin{assumption}
A1. Only single intersection scenarios are considered; A2.~A single lane is available per direction; A3.~The desired route of every agent is determined by a high-level route planning algorithm (Sec. \ref{sec:modeling_coordFrames});  A4. Agents are equipped with V2V communication; A5.~No communication failures or package dropouts occur; A6.~The MPC solutions at time $k$ are available to all agents at time $k+1$; A7.~Vehicle states are measurable and not subject to uncertainty. 
\end{assumption}
Assumptions A1, A2, A5 and A7 are common in the literature and are used to reduce complexity~\cite{Campos2014,Hult2016}. The use of a high-level planning algorithm which is postulated in A3 is quite common in AV architectures too \cite{Lim2018}. Lastly, \alex{A4 is necessary for a distributed control scheme and A6 can be satisfied by choosing the MPC sampling time appropriately}.

%------------------------------------------------------------------------
%------------------------------------------------------------------------
\subsection{Vehicle Kinematics}
\label{sec:modeling_kinematics}
\begin{figure}[t]	
	\hspace{5mm}
	\def\svgwidth{8.4cm}	
	\centering
    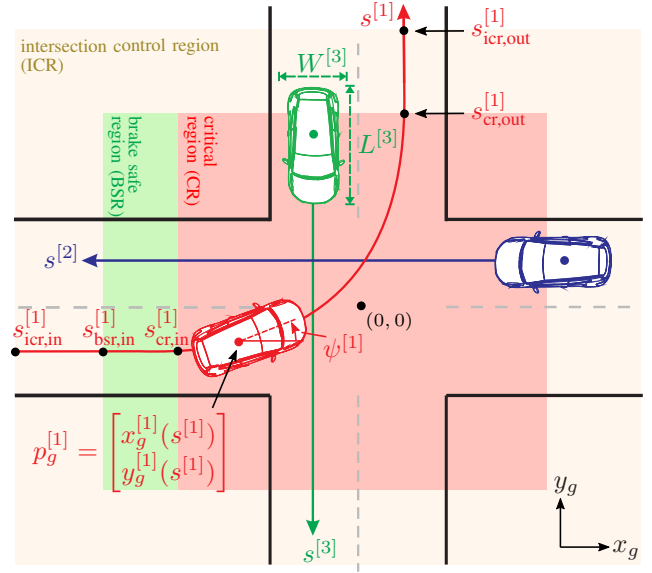
	\caption{Schematic of the intersection in the global frame, $(x_g, y_g)$, with its
	origin $(0,0)$ at the center of the intersection. For Agent 1, the respective 
	intersection regions are illustrated: inside ICR (beige), BSR (green), CR (red) 
	and outside ICR (white). The reference trajectory of Agent 1 is described by 
	a B-spline.}
%		Intersection coordination problem with respective regions for Agent 1.}
	%	\vspace*{-5mm}
	\label{fig:modeling_schematic}
	\vspace*{-6mm}
\end{figure}

%Let $N_A$ be the number of agents and 
%\(
%      \mathcal{A} 
%{}\triangleq{}
%      \{1, \ldots , N_A \}
%\).
\alex{Let 
\(
\mathcal{A} 
{}\triangleq{}
\{1, \ldots , N_A \}
\)
be the set of agents where $N_A$ is a positive integer.} It is a common approach in the literature to use Assumption A3 to define 
the kinematics of Agent $i \in \mathcal{A}$ along paths which are parametrized by a scalar 
$s^{[i]}$ as shown in \prettyref{fig:modeling_schematic} \cite{Hult2016,Molinari2018}. 
That said, the time evolution of velocity $v^{[i]}$ and path coordinate $s^{[i]}$ 
are described by a double integrator, while the drivetrain dynamics is modeled as 
a first-order system, yielding the following linear time-invariant state space 
representation
\begin{align}
\frac{d}{dt}
\begin{bmatrix}
{a}_{x}^{[i]} \\
{v}^{[i]} \\
{s}^{[i]}
\end{bmatrix} 
{}={} 
	\underbrace{\begin{bmatrix}
	-\frac{1}{T_{a_x}^{[i]}} & 0 & 0 \\
	1 & 0 & 0 \\
	0 & 1 & 0
	\end{bmatrix}}_{A^{[i]}}
	\underbrace{
	\vphantom{\begin{bmatrix}
		\frac{1}{T_{a_x}^{[i]}} \\
		0 \\
		0
		\end{bmatrix}}
	\begin{bmatrix}
	{a}_{x}^{[i]} \\
	{v}^{[i]} \\
	{s}^{[i]} 
	\end{bmatrix}}_{x^{[i]}} +
	\underbrace{\begin{bmatrix}
	\frac{1}{T_{a_x}^{[i]}} \\
	0 \\
	0
	\end{bmatrix}}_{B^{[i]}} \underbrace{\vphantom{\begin{bmatrix}
		\frac{1}{T_{a_x}^{[i]}} \\
		0 \\
		0
	\end{bmatrix}}
	a_{x,\text{ref}}^{[i]}}_{u^{[i]}},
\label{eq:modeling_kinematics_ssModelVehicle}
\end{align}
where $T_{a_{x}}^{[i]}$ denotes the dynamic drivetrain time constant and 
\(
      u^{[i]}
{}={}
      a_{x,\text{ref}}^{[i]}
\)
is the requested acceleration (acceleration set-point). 
We may derive the exact discretization of \prettyref{eq:modeling_kinematics_ssModelVehicle} 
using zero-order hold, that is, 
\(
      A_d^{[i]} 
{}={}
      e^{A^{[i]}T_s}
\)
and 
\(
      B_d^{[i]} 
{}={}
      \int_{0}^{T_s} e^{A^{[i]}\tau} B^{[i]} \mathrm{d}\tau
\)
where $T_s$ is the sampling time. 

%------------------------------------------------------------------------
%------------------------------------------------------------------------
\subsection{Local and Global Coordinate Frames}
\label{sec:modeling_coordFrames}
Equation \prettyref{eq:modeling_kinematics_ssModelVehicle} describes the motion 
of every agent with respect to their local coordinate frame along the path coordinate 
$s^{[i]}$. In \cite{Katriniok2017a}, collision avoidance is encoded in this local frame with respect to a joint collision point with another agent. This approach may not be suitable in multi-lane scenarios when a certain lateral distance to surrounding agents should be maintained. Then, a problem formulation in a global Cartesian frame appears to be more general and convenient. We, therefore, define a mapping function 
\(
      \mathcal{F}_p^{[i]}
{}:{}
      s^{[i]} 
{}\mapsto{}
      (x_g^{[i]}, y_g^{[i]}) 
%(= p_g^{[i]})
\), which maps the path coordinate $s^{[i]}$ of Agent $i$ to its global Cartesian coordinates. 
Functions $\mathcal{F}_p^{[i]}$ can be chosen to be B-splines \cite{deBoor1978}
\begin{align}
\begin{bmatrix}
x_g^{[i]}(s^{[i]}) \\ y_g^{[i]}(s^{[i]})
\end{bmatrix} = \mathcal{F}_p^{[i]}(s^{[i]}) \triangleq \begin{bmatrix}
\sum_{l=0}^{n_p}\alpha_{x,l}^{[i]}\,B_{x,l}^{[i],n}(s^{[i]}) \\
\sum_{l=0}^{n_p}\alpha_{y,l}^{[i]}\,B_{y,l}^{[i],n}(s^{[i]})
\end{bmatrix},
\end{align}
where $B_{x,l}^{[i],n}$ and $B_{y,l}^{[i],n}$ are B-spline basis polynomials of degree 
$n$ while $\alpha_{x,l}^{[i]}$ and $\alpha_{y,l}^{[i]}$ are spline coefficients.
We assume that the initial condition $s^{[i]}(t_0)=0$ holds.
For natural driving maneuvers, we use a spline degree of $n=3$. 
The number of spline coefficients, $n_p$, depends on the number of path 
points used for interpolating the agent's path \cite{deBoor1978}. 
According to Assumption A3, $\mathcal{F}_p^{[i]}$ is provided by a high-level 
route planning algorithm. 
Similarly, we assume that the heading angle $\psi^{[i]}(s^{[i]})$ and the path curvature 
$\kappa^{[i]}(s^{[i]})$ are provided as 
\(
      \mathcal{F}_\psi^{[i]}
{}:{} 
      s^{[i]} 
{}\mapsto{}
      \psi^{[i]}
\)
and 
\(
      \mathcal{F}_\kappa^{[i]}
{}:{} 
      s^{[i]} 
{}\mapsto{}
      \kappa^{[i]}
\)
respectively, either as separate spline curves or derived from the first and 
second derivative of $\mathcal{F}_p^{[i]}$ respectively. 

\subsection{Intersection Regions}
\label{sec:modeling_regions}
For agent coordination, we divide the area around the intersection in regions as 
shown in \prettyref{fig:modeling_schematic}. 
In the \textit{intersection control region} (ICR), that is, for 
\(
      s^{[i]}_{\text{icr,in}} 
{}\leq{}
      s^{[i]} 
{}<{}
      s^{[i]}_{\text{icr,out}}
\), 
the aim is to avoid collisions with crossing agents and to ensure rear-end collision avoidance at the same time. %while rear-end collision avoidance is independent of the ICR. 
The ICR can further be subdivided into the \textit{brake safe region} (BSR) and the 
\textit{critical region} (CR). 
In the BSR, \alex{defined as}
\(
      s^{[i]}_{\text{bs,in}} 
{}\leq{}
      s^{[i]} 
{}<{}
      s^{[i]}_{\text{cr,in}}
\), 
it is always safe to stop before entering the CR. Only in the CR, \alex{that is, for}
\(
s^{[i]}_{\text{cr,in}} 
{}\leq{}
s^{[i]} 
{}<{}
s^{[i]}_{\text{cr,out}}
\), collisions with crossing agents may happen and these need to be avoided by the control scheme. After leaving the CR and outside the ICR, only rear-end collisions must be prevented.
%where collisions with crossing agents need to be avoided.
%where the distributed control scheme needs to ensure collision-free \alex{trajectories}.

%-------------------------------------------------------------------------
%-------------------------------------------------------------------------
%-------------------------------------------------------------------------
% Intersection Motion Planning Problem
%-------------------------------------------------------------------------
%-------------------------------------------------------------------------
%-------------------------------------------------------------------------
%\vspace*{-4mm}
\section{Distributed Motion Planning Problem}
\label{sec:problem}

Following \cite{Katriniok2017a}, the motion planning problem is separable with 
respect to the agents' individual objectives and constraints while only collision 
avoidance couples the agents among each other. 
This way, we separate the subproblems using a primal decomposition technique.
%-------------------------------------------------------------------------
%-------------------------------------------------------------------------
\subsection{Local Agent Objectives and Constraints}
\label{sec:problem_localObj}
As local objectives, the motion planning regime should 1) maintain a desired speed 
(typically, close to the speed limit) and 2) enable efficient and comfortable 
driving. 
The former requirement translates into minimizing the deviation of the agent's speed,
$v^{[i]}$, from a reference speed $v_{\mathrm{ref}}^{[i]}$, while the latter is 
equivalent to reducing the demanded acceleration $u^{[i]}=a_{x,\text{ref}}^{[i]}$. 
Along a horizon of $N$ steps, we can encode these requirements in the following stage cost at time $k+j$ \alex{for $j=0,\ldots,N-1$}
\begin{align}
	\ell_{j}^{[i]}(x_{k+j\mid k}^{[i]},u_{k+j\mid k}^{[i]}) 
{}\triangleq{}& 
	q^{[i]} \, \bigl(v_{k+j\mid k}^{[i]} - v_{\mathrm{ref},k+j\mid k}^{[i]} \bigr)^2 
\notag 
\\
{}+{}&
	r^{[i]} \, ( u_{k+j\mid k}^{[i]} )^2  
\label{eq:problem_localObj_stageCost} 
\end{align}
and the terminal cost
\begin{align}
	\ell_{N}^{[i]}(x_{k+N\mid k}^{[i]}) 
{}\triangleq{} 
	q_N^{[i]} \, \bigl(v_{k+N\mid k}^{[i]} - v_{\mathrm{ref},k+N\mid k}^{[i]} \bigr)^2  
\label{eq:problem_localObj_termCost} 
\end{align}
where $q^{[i]} {}>{} 0$, $q_N^{[i]} {}>{} 0$ and $r^{[i]} {}>{} 0$ are positive weights. 
Besides local objectives, local agent constraints need to be accommodated as well. 
First, to account for actuator limitations, we constrain the demanded longitudinal 
acceleration by the input constraint
\begin{align}
	u_{k+j\mid k}^{[i]} 
{}\in{} 
	\mathcal{U}^{[i]} 
{}\triangleq{} 
	\left\{ 
	      u \in \mathbb{R} 
	  {}\mid{}
	      \underline{a}_{x}^{[i]} 
	    {}\leq{}
	      u 
	    {}\leq{}
	      \overline{a}_{x}^{[i]} 
	\right\}
\label{eq:input_constraints}
\end{align}
with some appropriate upper and lower bound $\overline{a}_{x}^{[i]}$ and 
$\underline{a}_{x}^{[i]}$, for $j=0,\ldots,N-1$. 

Moreover, every agent should account for the speed limit on the current 
road section while driving backwards is prohibited. 
We encode this requirement as a state constraint of the form 
\begin{align}
	x_{k+j\mid k}^{[i]} 
{}\in{}
	\mathcal{X}_{k+j\mid k}^{[i]} 
{}\triangleq{}
	\left\{ 
	      x 
	  {}\in{} 
	      \mathbb{R}^3 
	  {}\mid{}
		0 	
	      {}\leq{}
		[x]_2 
	      {}\leq{}
		\overline{v}_{k+j\mid k}^{[i]} 
	\right\}
\end{align}
with the upper bound $\overline{v}_{k+j\mid k}^{[i]}$, for $j=1,\ldots,N$. 

For turning agents, the lateral acceleration $a_y^{[i]}$, being equal to the product 
of the curvature $\kappa^{[i]}(s^{[i]})$ with the squared velocity $v^{[i]}$,
is bounded as follows
\begin{align}
	-\overline{a}_{y}^{[i]} 
{}\leq{}
	\kappa^{[i]}(s_{k+j\mid k}^{[i]}) 
{}\cdot{} 
	(v_{k+j\mid k}^{[i]})^2 
{}\leq{}
	\overline{a}_{y}^{[i]},
\label{eq:problem_localObj_alateral}
\end{align}
for $j=1,\ldots,N$, where $\overline{a}_{y}^{[i]}$ is an appropriately chosen 
upper bound. 
Additionally, it needs to be ensured that the total acceleration does not 
exceed the friction circle \cite{Rajamani2012}, that is, 
\begin{align}
    (a_{x,k+j\mid k}^{[i]})^2 + \left( \kappa^{[i]}(s_{k+j\mid k}^{[i]}) 
{}\cdot{} 
    (v_{k+j\mid k}^{[i]}\,)^2 \right)^2 
{}\leq{} 
    (\overline{a}_{\text{tot}}^{[i]})^2
\label{eq:problem_localObj_ahor}
\end{align}
for $j=1,\ldots,N$ and an appropriate upper bound $\overline{a}_{\text{tot}}^{[i]}$.

%-------------------------------------------------------------------------
%-------------------------------------------------------------------------
\subsection{Collision Avoidance}
\label{sec:problem_CA}

\subsubsection{Agent Conflict Sets}
\label{sec:problem_CA_conflictSets}
While the constraints in \prettyref{sec:problem_localObj} refer to the individual agent, CA constraints % are the \textit{complicating constraints} --- they 
couple the agents among each other. 
%In order to fully decouple every pair of conflicting agents, we impose CA constraints only on one of the agents. 

\alex{For crossing agents, we introduce} time-invariant priorities on the agents that are determined once and held constant during the maneuver \cite{Katriniok2017a}. We define the bijective prioritization function 
\(
\gamma
{}:{}
\mathcal{A} 
{}\rightarrow{}
\mathcal{A}
\), 
which assigns a unique priority to every agent --- a lower value corresponds to a higher priority. This way, we specify the set of agents $l {}\in{} \mathcal{A}$ which can collide with Agent $i$, but have a higher priority 
$\gamma(l) {}<{} \gamma(i)$, that is,
\begin{align}
\mathcal{A}_{c,\gamma}^{[i]} 
{}\triangleq{}
\bigl\{  
l {}\in{} \mathcal{A} 
{}\mid{}
\gamma(l) 
{}<{}
\gamma(i) 
\bigr\}.
\label{eq:problem_CA_conflictSetPrio} 
\end{align}
We refer to \(\mathcal{A}_{c,\gamma}^{[i]}\) as the time-invariant \textit{prioritized conflict set}. Moreover, we denote $\bar{\mathcal{A}}_{c,\gamma,k}^{[i]} \subseteq \mathcal{A}_{c,\gamma}^{[i]}$ as the set of higher priority agents 
%$l \in \mathcal{A}_{c,\gamma}^{[i]}$ 
that have not yet left the CR at time $k$.

In case of rear-end collision avoidance, we define $\mathcal{A}_{c,\text{ahead}}^{[i]}$ as the time-invariant set that contains the agents that are always in the same lane and ahead of Agent $i$. 
Furthermore, $\bar{\mathcal{A}}_{c,\text{ahead},k}^{[i]}$ refers to the time-varying set of agents $l \in \mathcal{A}$ at time $k$ that have been crossing (i.e., $l \in \mathcal{A}_{c,\gamma}^{[i]}$ or $i \in \mathcal{A}_{c,\gamma}^{[l]}$) 
and are now in the same lane and ahead of Agent $i$.
%Furthermore, $\bar{\mathcal{A}}_{c,\text{ahead},k}^{[i]} \subseteq \mathcal{A}_{c,\gamma}^{[i]}$ refers to the time-varying set of agents $l \in \mathcal{A}$ at time $k$ that have been crossing %(i.e., $l \in \mathcal{A}_{c,\gamma}^{[i]}$ or $i \in \mathcal{A}_{c,\gamma}^{[l]}$ ) 
%and are now in the same lane and ahead of Agent $i$.

%To fully decouple every pair of conflicting agents, we impose CA constraints only on one of the agents. 
To fully decouple the agents, we impose CA constraints only on one of two conflicting agents. Therefore, we consider the following cases at time $k$: 
%To fully decouple the agents, we compute the conflict sets at time $k$ as follows: 
\textbf{a)} If \mbox{Agent $i$} is \textbf{inside} the \textbf{ICR} and has \textbf{not yet passed the CR}, 
%it needs to avoid collisions with crossing agents and with those driving ahead and in the same lane, 
it imposes CA constraints on crossings agents and those that are driving ahead, i.e., on agents
\(
l 
{}\in{}
\mathcal{A}_{c,k}^{[i]} 
{}={}
\mathcal{A}_{c,\gamma}^{[i]} 
{}\cup{}
\mathcal{A}_{c,\text{ahead}}^{[i]}
\); 
\textbf{b)} If Agent $i$ is \textbf{inside} the \textbf{ICR} and has \textbf{passed the CR}, it needs to avoid collisions with agents 
\(
l \in
\mathcal{A}_{c,k}^{[i]} 
{}={} 
\mathcal{A}_{c,\text{ahead}}^{[i]}
{}\cup{}
\bar{\mathcal{A}}_{c,\text{ahead},k}^{[i]}
{}\cup{}
\bar{\mathcal{A}}_{c,\gamma,k}^{[i]}
\), that is, with those driving ahead and with higher priority agents that may be in the same lane and behind Agent $i$ after leaving the CR;
%we need to avoid rear-end collisions with agents driving ahead
% and with high priority agents which have not yet left the CR. %Thus, for the constraint set holds: 
\textbf{c)} If Agent $i$ is \textbf{outside} the \textbf{ICR}, we only impose rear-end CA constraints on agents
\(
l 
{}\in{}
\mathcal{A}_{c,k}^{[i]} 
{}={} 
\mathcal{A}_{c,\text{ahead}}^{[i]}
{}\cup{}
\bar{\mathcal{A}}_{c,\text{ahead},k}^{[i]}
\) 
that drive in the same lane and ahead of Agent $i$.

\subsubsection{Safety Regions and CA Constraint Formulation}
\begin{figure}[b]
	\centering	
%	\vspace*{3mm}
	\setlength\fwidth{0.44\textwidth}	
	\hspace*{-6mm}
 	% This file was created by matlab2tikz.
%
%The latest updates can be retrieved from
%  http://www.mathworks.com/matlabcentral/fileexchange/22022-matlab2tikz-matlab2tikz
%where you can also make suggestions and rate matlab2tikz.
%
\definecolor{mycolor1}{rgb}{0.00000,0.44700,0.74100}%
\definecolor{mycolor2}{rgb}{0.49400,0.18400,0.55600}%
\begin{tikzpicture}

\begin{axis}[%
width=0.951\fwidth,
height=0.609375\fwidth,%0.65625\fwidth, %0.75
at={(0\fwidth,0\fwidth)},
scale only axis,
%hide axis,
xmin=-6.61419992551159,
xmax=13.6720986429956,
ymin=-8,
ymax=5, %6, %8
axis background/.style={fill=white},
xmajorgrids,
ymajorgrids,
ytick={-8,-6,-4,-2,0,2,4,6,8},
yticklabels={-8,-6,-4,-2,0,2,4,6,8},
legend style={legend cell align=left, align=left, draw=white!15!black}
]
%\addplot [color=mycolor1, draw=none, mark size=5.0pt, mark=o, mark options={solid, mycolor1}]
%  table[row sep=crcr]{%
%0	0\\
%};
\addplot [color=mycolor1, draw=none, mark size=2.2pt, mark=*, mark options={solid, black}]
table[row sep=crcr]{%
	0	0\\
};
%\addlegendentry{data1}

\addplot[area legend, draw=black, fill=blue, fill opacity=0.5]
table[row sep=crcr] {%
x	y\\
-2.5	-1\\
2.5	-1\\
2.5	1\\
-2.5	1\\
}--cycle;
%\addlegendentry{data2}

\addplot[area legend, draw=black, fill=blue, fill opacity=0.2]
table[row sep=crcr] {%
x	y\\
2.5	0\\
1.5	1\\
1.5	-1\\
}--cycle;
%\addlegendentry{data3}

\addplot[area legend, draw=white, fill=blue, fill opacity=0.3]
table[row sep=crcr] {%
x	y\\
-4.5	-2\\%-3.92836282905962\\
5.5	-2\\%-3.92836282905962\\
5.5	2\\
-4.5	2\\
}--cycle;
%\addlegendentry{data4}

\addplot[area legend, draw=white, fill=blue, fill opacity=0.1]
table[row sep=crcr] {%
x	y\\
-5.5	-4\\
8.5	-4\\
8.5	2\\
-5.5	2\\
}--cycle;
%\addlegendentry{data5}

\addplot [color=blue, draw=none, mark size=2.0pt, mark=*, mark options={solid, blue}]
  table[row sep=crcr]{%
-5.5	-4\\
};
%\addlegendentry{data6}

\addplot [color=blue, draw=none, mark size=2.0pt, mark=*, mark options={solid, blue}]
  table[row sep=crcr]{%
8.5	2\\
};
%\addlegendentry{data7}

%\addplot [color=mycolor2, draw=none, mark size=5.0pt, mark=o, mark options={solid, mycolor2}]
%  table[row sep=crcr]{%
%9	-4\\
%};
\addplot [color=mycolor2, draw=none, mark size=2.2pt, mark=*, mark options={solid, black}]
table[row sep=crcr]{%
	9.5	-4.5\\
};
%\addlegendentry{data8}

\addplot[area legend, draw=black, fill=red, fill opacity=0.5]
table[row sep=crcr] {%
x	y\\
8.22767650188909	-6.87301346733533\\
12.057898717484	-3.65907541890263\\
10.7723234981109	-2.12698653266467\\
6.94210128251602	-5.34092458109737\\
}--cycle;
%\addlegendentry{data9}

\addplot[area legend, draw=black, fill=red, fill opacity=0.2]
table[row sep=crcr] {%
x	y\\
11.4151111077974	-2.89303097578365\\
10.00627905499193	-2.76977414235121\\
11.291854274365	-4.30186302858917\\
}--cycle;
%\addlegendentry{data10}

\addplot[area legend, draw=black, fill=red, fill opacity=0.2]
table[row sep=crcr] {%
x	y\\
6.94210128251602	-6.87301346733533\\
12.057898717484	-6.87301346733533\\
12.057898717484	-2.12698653266467\\
6.94210128251602	-2.12698653266467\\
}--cycle;

\tikzset{
	schraffiert/.style={pattern=north west lines,pattern color=#1},
	schraffiert/.default=black
}

\addplot[schraffiert=blue, draw=none]%, draw=none, fill=red, fill opacity=0.2]
table[row sep=crcr] {%
	x	y\\
	6.94210128251602	-4\\
	8.5	                -4\\
	8.5  	            -2.12698653266467\\
	6.94210128251602	-2.12698653266467\\
}--cycle;

%\addlegendentry{data11}

\addplot [color=red, draw=none, mark size=2.0pt, mark=*, mark options={solid, red}]
  table[row sep=crcr]{%
6.94210128251602	-6.87301346733533\\
};
%\addlegendentry{data12}

\addplot [color=red, draw=none, mark size=2.0pt, mark=*, mark options={solid, red}]
  table[row sep=crcr]{%
12.057898717484	-2.12698653266467\\
};
%\addlegendentry{data13}

\end{axis}

\begin{axis}[%
width=1.227\fwidth,
height=0.609\fwidth,
at={(-0.16\fwidth,-0.101\fwidth)},
scale only axis,
xmin=0,
xmax=1,
ymin=0,
ymax=1,
axis line style={draw=none},
ticks=none,
axis x line*=bottom,
axis y line*=left,
legend style={legend cell align=left, align=left, draw=white!15!black}
]
\end{axis}

%\draw[color=blue,arrows = -{Stealth[inset=0pt]}] (2.43,2.93) -- (4.93,2.93);
\draw[thick,arrows = -{Stealth[inset=0pt]}] (2.43,2.93) -- (3.93,2.93);
\draw[thick,arrows = -{Stealth[inset=0pt]}] (2.43,2.93) -- (2.43,4.43);
\draw[color=red,arrows = -{Stealth[inset=0pt]}] (5.91,1.29) -- (7.21,2.35);
\draw[color=blue,arrows = -] (4.8,1.3) -- (5.05,1.55);
\node[right, align=left, font=\color{blue}] at (0.2,1.10) {$\underline{p}^{[i]}_{}$};
\node[right, align=left, font=\color{blue}] at (5.0,4.0) {$\overline{p}^{[i]}_{}$};
\node[right, align=left, font=\color{red}] at (4.3,0.45) {$\underline{p}^{[l]}_{}$};
\node[right, align=left, font=\color{red}] at (6.4,2.5) {$\overline{p}^{[l]}_{}$};
\node[right, align=left, font=\color{red}] at (6.8,1.95) {$n_\psi^{[l]}$};
\node[right, align=left, font=\color{blue}] at (4.1,1.25) {$A^{i,l}$};
%\node[right, align=left, font=\color{red}] at (6.3,0.2) {\small\textbf{RV}};
%\node[right, align=left, font=\color{blue}] at (0.7,3.9) {\small\textbf{HV}};

\node[right, align=left, font=\color{black}] at (2.5,4.4) {$y^{[i]}$};
\node[right, align=left, font=\color{black}] at (3.55,2.6) {$x^{[i]}$};
%\node[right, align=left, font=\color{blue}] at (4.9,3.0) {$v^{[i]}$};
\node[right, align=left, font=\color{black}] at (1.0,1.75) {\scriptsize\textbf{motion dep. safety region}};
\node[right, align=left, font=\color{white}] at (1.35,2.35) {\scriptsize\textbf{basic safety region}};

\draw[color=blue,|-|] (4.97, 2.30) --  (5.53, 2.30);
\draw[color=blue,|-|] (4.80, 1.48)  -- (4.80, 2.16);
\node[right, align=left, font=\color{blue}] at (4.87,2.55) {${L}^{i,l}_{}$};
\node[right, align=left, font=\color{blue}] at (3.95,1.75) {${W}^{i,l}_{}$};
\end{tikzpicture}%
	\vspace*{-7mm}
	\caption{Agent $i$'s safety region (basic and motion dependent) 
	along with Agent $l$'s bounding box in Agent $i$'s Cartesian body 
	frame. $A^{i,l}$ represents their overlap.}
%		Overlap of agent $i$'s safety region and agent $l$'s bounding box.}
%		\vspace*{-9mm}
	\label{fig:problem_areaOverlap}
	%	\vspace*{-9mm}
\end{figure}
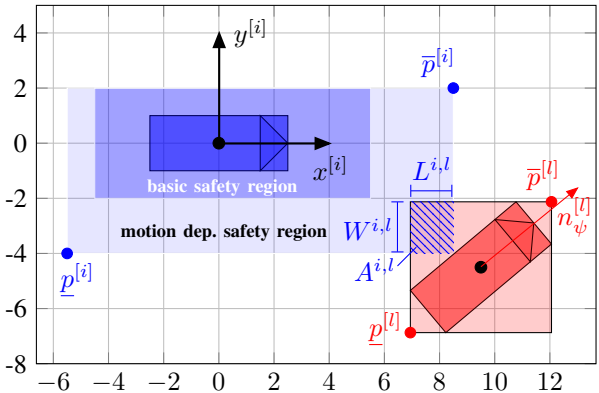

For each agent $i \in \mathcal{A}$, we define a safety region around the vehicle 
for every $l \in \mathcal{A}_{c,k}^{[i]}$, which must not be intersected by 
the bounding box of the conflicting Agent $l$. We denote by
\begin{align}
n_\psi^{[l]} = R_\psi^{[i], \T} R_\psi^{[l]} \begin{bmatrix}
1 & 0
\end{bmatrix}^{\T}
\end{align}
the unit vector pointing in the direction of motion of \mbox{Agent $l$} in the 
Cartesian body frame of Agent $i$ where $R_\psi^{[i]}$ and $R_\psi^{[l]}$ 
are the respective rotation matrices --- see \prettyref{fig:problem_areaOverlap}. 
%We then define the safety region in the local $x$-direction in front of (xf) 
%and behind (xr) \alex{Agent} $i$ as 
We then define the safety region in the \alex{longitudinal} direction in front of (xf) and behind (xr) \alex{Agent} $i$ as 
\begin{subequations}
\begin{align}
      d_{\text{safe,xf}}^{[i]} 
{}\triangleq{}& 
      \tilde{d}_{\text{safe,xf}}^{[i]} 
{}+{} 
      v^{[i]} t_{\text{gap},x}^{[i]}, 
\\
      d_{\text{safe,xr}}^{[i]} 
{}\triangleq{}& 
      \tilde{d}_{\text{safe,xr}}^{[i]}
{}+{} 
      v^{[i]} t_{\text{gap},x}^{[i]} 
{}\cdot{} 
      \max \big\{ 0, [1~0] n_\psi^{[l]} \big\}\label{eq:d_safe_xr},
\end{align}
\end{subequations}
where $\tilde{d}_{\text{safe,xf}}^{[i]}$ and $\tilde{d}_{\text{safe,xr}}^{[i]}$
are basic safety distances which are independent of the agents' motion. 
In addition, we want the agents to keep a velocity-dependent safety distance where $t_{\text{gap},x}$ is the respective time gap. 
\alex{The rear safety distance $d_{\text{safe,xr}}^{[i]}$ should only be increased if \alex{Agent} $l$ is driving in the same direction as \alex{Agent} $i$, and not if it is crossing.}
%For $d_{\text{safe,xr}}^{[i]}$, it is reasonable to let this distance depend 
%on the heading of agent $l$, especially if agent $l$ is turning into agent 
%$i$'s lane. 
That said, in \eqref{eq:d_safe_xr}, $[1~0] n_\psi^{[l]}$ is the
projection of Agent $l$'s heading vector onto Agent $i$'s body frame $x$-axis.

%In the $y$-direction, i.e., left (yl) and right (yr) from Agent $i$, we consider the basic safety distances 
In the \alex{lateral} direction, i.e., left (yl) and right (yr) from Agent $i$, we consider the basic safety distances 
\(
	\tilde{d}_{\text{safe},\text{yl}}^{[i]}
\)
and 
\(
	\tilde{d}_{\text{safe},\text{yr}}^{[i]}
\)
which are independent of the agents' motion. 
If Agent $l$ is moving from the side towards Agent $i$, the safety distance increases
depending on the speed and heading of Agent $l$, that is,
\begin{align}
	d_{\text{safe,\text{yl}/\text{yr}}}^{[i]} 
{}\triangleq{}
	\tilde{d}_{\text{safe,\text{yl}/\text{yr}}}^{[i]} 
{}+{} 
	v^{[l]} t_{\text{gap},y}^{[i]} 
{}\cdot{} 
	\max \bigl\{ 0, \mp [0~1] n_\psi^{[l]} \bigr\},
\end{align}
where $t_{\text{gap},y}$ is an appropriate time gap, and $[0~1] n_\psi^{[l]}$ is the 
projection of Agent $l$'s heading vector onto Agent $i$'s body frame $y$-axis. 
\alex{Applying this formulation,} the safety distance to Agent $l$ increases, even as it moves away from Agent $i$. \alex{In this case, though, the problem solution is not affected.}
%The attentive reader might have recognized that the safety distance to Agent $l$ increases, even as it moves away from Agent $i$. 
%However, as the safety distance is enlarged in the opposite direction of Agent $l$'s motion, the problem solution is not affected.
%The interaction of both agents or even with other agents, though, is not affected.
%\alex{
%\begin{remark}
%	Motion dependent safety distances, especially in the lateral direction, are required to avoid an infeasible OCP on narrow roads with oncoming traffic.
%%	A constant safety distance around each agent might lead to an infeasible OCP in case 
%%	of oncoming vehicles in narrow roads. 
%%	Using motion dependent safety distances actually prevents such situations.
%\end{remark}}

We now define the lower-left and upper-right corner points,
\(\underline{p}^{[i]}\) and \(\overline{p}^{[i]}\) of Agent $i$'s safety region 
(see \prettyref{fig:problem_areaOverlap}) in its Cartesian body frame as 
%\vspace*{-1mm}
\begin{align}
	\underline{p}_{}^{[i]} 
{}\triangleq{}
	\begin{bmatrix}
	    -\frac{L^{[i]}}{2} - d_{\text{safe,xr}}^{[i]} 
	\\
	    -\frac{W^{[i]}}{2} - d_{\text{safe,\text{yr}}}^{[i]}
	\end{bmatrix},~
	\overline{p}_{}^{[i]} 
{}\triangleq{} 
	\begin{bmatrix}
	    \frac{L^{[i]}}{2} + d_{\text{safe,xf}}^{[i]} 
	\\
	    \frac{W^{[i]}}{2} + d_{\text{safe,\text{yl}}}^{[i]}
	\end{bmatrix},
\end{align}
where $L^{[i]}$ and $W^{[i]}$ are the length and width of Agent $i$, respectively. 
To determine collisions between agents, we examine the area overlap 
of Agent $i$'s safety region and Agent $l$'s bounding box. 
If there is no overlap, both agents are safe. 
If the edges of both rectangles are not perpendicular to each other, a closed-form expression 
is hard to determine. For that reason, we resort to over-approximating Agent $l$ with a bounding box
whenever the rectangle edges are not perpendicular to each other as shown in
\prettyref{fig:problem_areaOverlap}. 
We denote the lower right and upper left corner point of the over-approximated rectangle in Agent $i$'s reference frame as $\underline{p}^{[l]}$ and $\overline{p}^{[l]}$. \alex{Then, the length $L^{i,l}$ and width $W^{i,l}$ of the overlapping area can easily be obtained as follows}
%These can easily be obtained by computing the minimum and maximum coordinates of the original bounding box as follows
\begin{subequations}
\begin{align}
	L^{i,l} 
{}\triangleq{}&
	\min 
	  \bigl\{ 
	    [\overline{p}^{[i]}]_1, [\overline{p}^{[l]}]_1  
	  \bigr\} 
{}-{}
	\max 
	  \bigl\{ 
	    [\underline{p}^{[i]}]_1, [\underline{p}^{[l]}]_1  
	  \bigr\}, 
\\
	W^{i,l} 
{}\triangleq{} &
	\min 
	  \bigl\{
	    [\overline{p}^{[i]}]_2, [\overline{p}^{[l]}]_2 
	  \bigr\} 
{}-{}
	\max 
	  \bigl\{ 
	    [\underline{p}^{[i]}]_2, [\underline{p}^{[l]}]_2 
	  \bigr\}.
\end{align}
\end{subequations}
If there is no overlap, $L^{i,l}$ and/or $W^{i,l}$ are less than zero. 
Therefore, the overlap of Agent $i$'s safety region and Agent $l$'s bounding box is
\begin{align}
	A^{i,l} 
{}\triangleq{}
	\max 
	    \bigl\{0, L^{i,l} \bigr\} 
{}\cdot{}
	\max \bigl\{ 0, W^{i,l} \bigr\}.
\end{align}%
Finally, the following equality constraint needs to be satisfied to guarantee collision avoidance \alex{for every agent in the conflict set \(\mathcal{A}_{c,k}^{[i]}\) (see \prettyref{sec:problem_CA}) and} for every time step $k+j$, $j=1,\ldots,N$ over the prediction horizon:
%Finally, the following equality constraint needs to be satisfied to guarantee collision 
%avoidance for every time step $k+j$, $j=1,\ldots,N$ over the prediction horizon:
\begin{align}
A^{i,l}_{k+j\mid k} = 0,~\forall l \in \mathcal{A}_{c,k}^{[i]}.
\end{align}
%holds.
\vspace*{-3mm}
\begin{remark}
Over-approximating Agent $l$'s bounding box does not entail any undesired conservatism. 
A tighter approximation, though, might be preferable for future use cases 
such as multi-lane scenarios. 	
\end{remark}
%\begin{remark}
%A constant safety distance around each agent might lead to an infeasible OCP in case 
%of oncoming vehicles in narrow roads. 
%Using motion dependent safety distances actually prevents such situations.
%\end{remark}

%-------------------------------------------------------------------------
%-------------------------------------------------------------------------
\subsection{Minimum Spatial Preview}
\label{sec:problem_minPreview}

To guarantee collision avoidance in the distributed setting, the spatial preview of every agent $i$ has to be of sufficient length while crossing the intersection, see \cite{Katriniok2017a}. 
%To guarantee local and global convergence of the distributed control scheme,
%we pursue a similar approach as in \cite{Katriniok2017a}. 
Particularly, if Agent $i$ has approached its BSR, i.e., 
\(
      s_{\text{bsr,in}}^{[i]}
{}\leq{}
      s_k^{[i]} 
{}<{} 
      s_{\text{cr,in}}^{[i]}
\), 
it is safe to stop before entering the CR where collision can happen. 
If there are still crossing agents with higher priority that have not yet left 
the CR, that is, if
\(
%      \mathcal{A}_{c,\gamma}^{[i]}
      \bar{\mathcal{A}}_{c,\gamma,k}^{[i]}
{}\neq{} 
      \emptyset
\), 
the preview of Agent $i$ has at least to cover the agent's CR \alex{to avoid unforeseen conflicts with other agents}. 
This requirement is equivalent to the constraint that Agent $i$ leaves the 
CR \alex{at the latest} at the final time step $k+N$ of the prediction horizon, i.e., 
\(
      s_{k+N\mid k}^{[i]} 
{}\geq{} 
      s_{\text{cr,out}}^{[i]}
\). 
If this constraint cannot be satisfied, Agent $i$ should stop before entering the CR. 
In essence, we need to encode the \textit{conditional constraint}
\begin{align}
      \mathrm{\mathbf{IF~NOT}}~ 
	      s_{k+N\mid k}^{[i]} 
	{}\geq{} 
	      s_{\text{cr,out}}^{[i]} 
      ~ 
      \mathrm{\mathbf{THEN}}~
	      s_{k+N\mid k}^{[i]} 
	{}\leq{} 
	      s_{\text{stop}}^{[i]},
\label{eq:problem_minPreview_stopCondCons}
\end{align}
where 
\(
      s_{\text{stop}}^{[i]} 
{}<{} 
      s_{\text{cr,in}}^{[i]}
\) 
refers to the stop line in the BSR. 
If all agents with higher priority have already left the critical region, we set 
\(s_{\text{cr,out}}^{[i]}\) to zero, thus allowing Agent $i$ to finally pass the 
critical region, referred to as \text{liveness} of the control scheme \cite{Kim2014}. 
To encode conditional constraints of the form
\begin{align}
      \mathrm{\mathbf{IF~NOT}}~ 
	    g(x) 
	 {}\geq{}
	    0~ 
      \mathrm{\mathbf{THEN}}~
	    h(x) 
	 {}\leq{}
	    0, 
\label{eq:problem_minPreview_condCons}
\end{align}%\\[-6mm]
with the constraint functions $g: \mathbb{R}^n \rightarrow \mathbb{R}$ and 
$h: \mathbb{R}^n \rightarrow \mathbb{R}$, we can write  \prettyref{eq:problem_minPreview_condCons} equivalently as
\mbox{\(
    \lnot ( g(x) {}\geq{} 0) 
{}\Rightarrow{}
    (h(x) \leq 0),
\)}
or, what is the same
\begin{align}
(-g(x)\leq 0) \lor (h(x) \leq 0)
\label{eq:problem_minPreview_logicCons}
\end{align}
where $\lnot$ denotes the logical NOT operator. 
By using the plus operator $[{}\cdot{}]_{\pplus}$, condition 
\prettyref{eq:problem_minPreview_logicCons} can be cast as an equality 
constraint of the form
\begin{align}
      [-g(x)]_{\pplus} {}\cdot{} [h(x)]_{\pplus} 
{}={} 
      0.
\label{eq:plus_op_reformulation}
\end{align}%\\[-6mm]
By virtue of \prettyref{eq:plus_op_reformulation}, 
condition \prettyref{eq:problem_minPreview_stopCondCons} can then be rewritten as
\begin{align}
      \bigl[ -s_{k+N\mid k}^{[i]} + s_{\text{cr,out}}^{[i]} \bigr]_{\pplus} 
{}\cdot{}  
      \bigl[ s^{[i]}_{k+N\mid k} - s_{\text{stop}}^{[i]} \bigr]_{\pplus} 
{}={} 
      0.
\end{align}

%-------------------------------------------------------------------------
%-------------------------------------------------------------------------
\subsection{Optimal Control Problem}
\label{sec:problem_OCP}
Based on the optimized position, velocity and heading trajectories 
\(
      (x_{g,\cdot\mid k-1}^{[l]},
      y_{g,\cdot\mid k-1}^{[l]},
      \psi_{\cdot\mid k-1}^{[l]},
      v_{\cdot \mid k-1}^{[l]})
\) 
of conflicting agents $l \in \mathcal{A}_{c,k}^{[i]}$ that have been transmitted 
at time $k-1$ (Assumption A6), every agent $i \in \mathcal{A}$ solves the following OCP 
at time $k$
\begin{subequations}
	\label{eq:problem_OCP_OCPstatement}
\begin{align}
&\hspace*{-7mm}
    \mathmin_{ \{u_{k+j\mid k}\}_{j=0}^{N-1} }\ 
      \ell_{N}^{[i]}(x_{k+N\mid k}^{[i]}) 
    {}+{} 
      \sum_{j=0}^{N-1} \ell_{j}^{[i]}(x_{k+j\mid k}^{[i]},u_{k+j\mid k}^{[i]}) 
\label{eq:problem_OCP_OCPstatement_cost}
\\
      \mathst~&~ x_{k+j+1\mid k}^{[i]} 
{}={} 
      A_d^{[i]} x_{k+j\mid k}^{[i]} {}+{} B_d^{[i]} u_{k+j\mid k}^{[i]} 
\label{eq:problem_OCP_OCPstatement_dynamics}  
\\
&~ 
      u_{k+j \mid k}^{[i]} 
{}\in{} 
      \mathcal{U}^{[i]},\,~~~~~~~~~~~~~~j=0,\ldots,N-1 
\label{eq:problem_OCP_OCPstatement_inputs}
\\[0.5mm]
&~ 
      x_{k+j \mid k}^{[i]} 
{}\in{} 
      \mathcal{X}_{k+j\mid k}^{[i]},\,~~~~~~~~~~j=1,\ldots,N 
\label{eq:problem_OCP_OCPstatement_states}
\\[0.5mm]
&~ 
      -\overline{a}_{y}^{[i]} 
{}\leq{}
      a_{y,k+j\mid k}^{[i]} 
{}\leq{}
      \overline{a}_{y}^{[i]},~~j=1,\ldots,N  
\label{eq:problem_OCP_OCPstatement_ay}
\\[0.5mm]
&~ 
      (a_{\text{tot},k+j\mid k}^{[i]})^2 
{}\leq{} 
      (\overline{a}_{\text{tot}}^{[i]})^2,~~~~~j=1,\ldots,N  
\label{eq:problem_OCP_OCPstatement_ahor}
\\[0.5mm]
&~ 
      A^{i,l}_{k+j\mid k} {}={} 0,~ \forall l \in \mathcal{A}_{c,k}^{[i]},~~~j=1,\ldots,N  
\label{eq:problem_OCP_OCPstatement_CA}
\\[0.5mm]
&~
      \bigl[ -s_{k+N\mid k}^{[i]} {}+{} s_{\text{cr,out}}^{[i]} \bigr]_{\pplus} \hspace*{-1mm} 
{}\cdot{} 
      \bigl[ s^{[i]}_{k+N\mid k} - s_{\text{stop}}^{[i]} \bigr]_{\pplus} \hspace*{-1mm}
{}={} 
      0, 
\label{eq:problem_OCP_OCPstatement_minMeanVel}
\\[0.5mm]
&~ 
      x_{k\mid k}^{[i]} 
{}={} 
      x_{k}^{[i]},
\end{align}
\end{subequations}
where 
%\mbox{\(
%	a_{\text{hor},k+j\mid k}^{[i]} 
%{}={} 
%	[
%	  (a_{x,k+j\mid k}^{[i]})^2 + (\kappa^{[i]}(s_{k+j\mid k}^{[i]}) 
%      {}\cdot{} 
%	  (v_{k+j\mid k}^{[i]})^2)^2 
%	]^{\nicefrac{1}{2}}
%\)}
\(
	a_{\text{tot},k+j\mid k}^{[i]} 
	{}={} 
	[
	(a_{x,k+j\mid k}^{[i]})^2 + (a_{y,k+j\mid k}^{[i]})^2 
	]^{\nicefrac{1}{2}}
\)
refers to the total acceleration in \prettyref{eq:problem_localObj_ahor} and 
\mbox{\(
  	a_{y,k+j\mid k}^{[i]}
{}={} 
  	\kappa^{[i]}(s_{k+j\mid k}^{[i]}) 
{}\cdot{} 
  	(v_{k+j\mid k}^{[i]})^2
\)}
to the lateral acceleration in \eqref{eq:problem_localObj_alateral}. 
At every time instant $k$, Agent $i$ solves the above optimization problem 
and obtains an optimal sequence 
\(
      (u_{k \mid k}^{[i],\star},\ldots,u_{k+N-1 \mid k}^{[i],\star})
\), 
the first element of which, \(u_{k\mid k}^{[i],\star}\), is applied to the plant. 
Moreover, the agents' optimized trajectories 
\(
      (x_{g,\cdot\mid k}^{[i]\star},
      y_{g,\cdot\mid k}^{[i]\star},
      \psi_{\cdot\mid k}^{[i]\star},
      v_{\cdot \mid k}^{[i]\star})
\) 
are transmitted to the other agents via V2V communication. 

Due to \eqref{eq:problem_OCP_OCPstatement_ahor}, \eqref{eq:problem_OCP_OCPstatement_CA} and \eqref{eq:problem_OCP_OCPstatement_minMeanVel}, Problem \prettyref{eq:problem_OCP_OCPstatement} 
is nonconvex. It can, however, be reformulated in a form so that it can be solved efficiently 
in real time using PANOC.%\cite{Sathya2018,Stella2017}.

%-------------------------------------------------------------------------
%-------------------------------------------------------------------------
%-------------------------------------------------------------------------
% Fast Numerical Solution of the NMPC Problem
%-------------------------------------------------------------------------
%-------------------------------------------------------------------------
%-------------------------------------------------------------------------
\section{Fast Numerical NMPC Solution}% of the\\NMPC Problem}
\label{sec:numsol}
The proximal averaged Newton-type method for optimal control (PANOC) has been proposed 
by the authors in \cite{Stella2017} (see also~\cite{Sathya2018,hermans2018penalty} 
for applications of PANOC to obstacle avoidance MPC). 
PANOC is a first-order method that combines projected gradient iterations with 
quasi-Newtonian directions for fast convergence. The key to guaranteeing global 
convergence of PANOC is the forward-backward envelope (FBE) introduced 
in~\cite{patrinos2013proximal} for convex problems and further extended to nonconvex problems in~\cite{stella2016forward,themelis2016forward}.
%----------------------------------------------------------------------------

To solve problem \prettyref{eq:problem_OCP_OCPstatement} for \alex{Agent} \(i\) with PANOC, 
it is reformulated and cast in the form
\begin{align}
  \mathmin_{
	      {u}_{\cdot\mid k}^{[i]} \in U_k^{[i]}
           } 
	    \phi^{[i]}_k({u}_{\cdot\mid k}^{[i]}; z^{[i]}_k),
\label{reformulated_problem}
\end{align}
where ${u}_{\cdot\mid k}^{[i]} = [u_{k\mid k}^{[i]},\ldots,u_{k+N-1\mid k}^{[i]}]^{\T}$
is the vector of predicted control actions of Agent $i$,  and
\(
	z_k^{[i]} 
{}={} 
	[
	    x_{k}^{[i],{\T}},
	    (x_{g,\cdot\mid k-1}^{[l],{\T}},
	    y_{g,\cdot\mid k-1}^{[l],{\T}},
	    \psi_{\cdot\mid k-1}^{[l],{\T}},
	    v_{\cdot \mid k-1}^{[l],{\T}}
	    )_{l\in\mathcal{A}_{c,k}^{[i]}}
	]^{\T}
\)
is a parameter vector which provides to Agent \(i\) all necessary measured 
information. PANOC requires that $\phi^{[i]}_k$ are continuously differentiable 
functions in \({u}_{\cdot\mid k}^{[i]}\) with Lipschitz-continuous gradient and 
that sets \(U_k^{[i]}\) are closed and such that we can easily compute projections thereon.

To that end, we first eliminate the state sequence using \eqref{eq:problem_OCP_OCPstatement_dynamics},
that is, we substitute 
\vspace*{-1mm}
\begin{align*}
      x_{k+j\mid k}^{[i]}( {u}_{\cdot\mid k}^{[i]}  )
{}={} 
      (A_d^{[i]})^{j} x_{k\mid k}^{[i]} 
{}+{} 
      \sum_{\iota{}={}0}^{j-1} (A_d^{[i]})^{j-1-\iota}B_{d}^{[i]}u_{k+\iota\mid k}^{[i]}.
\end{align*}\\[-3mm]
For the input constraints \eqref{eq:problem_OCP_OCPstatement_inputs} we define the set 
\(
      U_{k}^{[i]} 
{}\triangleq{} 
      {\{ {u}_{\cdot\mid k}^{[i]}
	     {}\mid{} u_{k+j \mid k}^{[i]} 
      {}\in{} 
	    \mathcal{U}^{[i]}, j=0,\ldots,N-1 
      \}}
\), 
which is a rectangle, because of the definition of \(\mathcal{U}^{[i]}\) 
in \eqref{eq:input_constraints}. The remaining equality and inequality constraints, namely 
\eqref{eq:problem_OCP_OCPstatement_states} to \eqref{eq:problem_OCP_OCPstatement_minMeanVel},
are modeled as \textit{soft constraints} and the \alex{quadratic} penalty method \cite[Chap.~17]{Nocedal2006} is used to ensure their satisfaction. 

In particular, equality 
constraints \eqref{eq:problem_OCP_OCPstatement_CA} and \eqref{eq:problem_OCP_OCPstatement_minMeanVel}
can be concisely written as 
\(
      h_{s}({u}_{\cdot\mid k}^{[i]}) 
{}={} 
      0
\) 
for \(s=1,\ldots,n_{\mathrm{eq}}\).
We introduce the associated penalty functions
%\begin{equation}
\begin{align*}
      \psi_{\mathrm{eq},s}({u}_{\cdot\mid k}^{[i]}, \beta_{\mathrm{eq}, s})
{}\triangleq{}
      \beta_{\mathrm{eq}, s} 
{}\cdot{}
      h_{s}({u}_{\cdot\mid k}^{[i]})^2
\end{align*}
%\end{equation}
with positive weights \(\beta_{\mathrm{eq}, s}>0\) for \(s=1,\ldots,n_{\mathrm{eq}}\). 
Inequality constraints can be concisely stated as 
\(
 g_{s}({u}_{\cdot\mid k}^{[i]}) \leq 0,
\)
for \(s = 1,\ldots, n_{\mathrm{ineq}}\). 
For each such constraint, we define the penalty function 
\begin{align*}
      \psi_{\mathrm{ineq}, s}({u}_{\cdot\mid k}^{[i]}, \beta_{\mathrm{ineq}, s})
{}\triangleq{}
      \beta_{\mathrm{ineq}, s}
{}\cdot{}
      [{}g_{s}({u}_{\cdot\mid k}^{[i]}){}]_{\pplus}^2,
\end{align*}
with positive weights \(\beta_{\mathrm{ineq}, s} > 0\). 
This way, we define the modified cost 
\begin{multline}
    \phi^{[i]}_k({u}_{\cdot\mid k}^{[i]}; z^{[i]}_k, (\beta_{\mathrm{eq},s})_s, (\beta_{\mathrm{ineq},s})_s) 
\\  
{}\triangleq{}     
    \ell_{N}^{[i]}(x_{k+N\mid k}^{[i]}({u}_{\cdot\mid k}^{[i]})) 
{}+{} 
    \sum_{j=0}^{N-1} \ell_{j}^{[i]}(x_{k+j\mid k}^{[i]}({u}_{\cdot\mid k}^{[i]}),u_{k+j\mid k}^{[i]}) 
\\ 
{}+{}
    \sum_{s=1}^{n_{\mathrm{eq}}} \psi_{\mathrm{eq},s}({u}_{\cdot\mid k}^{[i]}, \beta_{\mathrm{eq}, s})
{}+{}   
    \sum_{s=1}^{n_{\mathrm{ineq}}} \psi_{\mathrm{ineq}, s}({u}_{\cdot\mid k}^{[i]}, \beta_{\mathrm{ineq}, s}).
\end{multline}
Now, the problem is in the form \eqref{reformulated_problem} and can be solved
using PANOC. Note that the constraints are satisfied if and only if 
\(
      \psi_{\mathrm{eq},s}({u}_{\cdot\mid k}^{[i]}, \beta_{\mathrm{eq}, s})
{}={}
      0
\) 
and 
\(
      \psi_{\mathrm{ineq}, s}({u}_{\cdot\mid k}^{[i]}, \beta_{\mathrm{ineq}, s})
{}={}
      0
\)
for some weight parameters and for all $s$. 
%In order for the constraints to be satisfied, we apply a \alex{quadratic} penalty method \cite{Nocedal2006}: we select a tolerance $\epsilon_{s}>0$ for the constraints and we require that $\psi_{\mathrm{eq},s}/\beta_{\mathrm{eq},s} < \epsilon_{s}$ for all $s$ (similarly, $\psi_{\mathrm{ineq},s}/\beta_{\mathrm{ineq},s} < \epsilon_{s}$ for the inequality constraints).
In order for the constraints to be satisfied, we select a tolerance $\epsilon_{s}>0$ for the constraints and we require that $\psi_{\mathrm{eq},s}/\beta_{\mathrm{eq},s} < \epsilon_{s}$ for all $s$ (similarly, $\psi_{\mathrm{ineq},s}/\beta_{\mathrm{ineq},s} < \epsilon_{s}$ for the inequality constraints). If this is not satisfied for some $s$, we \alex{multiply} the corresponding weight \alex{by 10} and solve the problem, warm-starting with the previous approximate solution.

%---------------------------------------------------------------------------
%---------------------------------------------------------------------------
%---------------------------------------------------------------------------
% SIMULATION RESULTS
%---------------------------------------------------------------------------
%---------------------------------------------------------------------------
%---------------------------------------------------------------------------
\section{Simulation Results}
\label{sec:results}
\begin{figure*}[t]
	\begin{center}
		\setlength\fwidth{0.72\textwidth}
 		\input{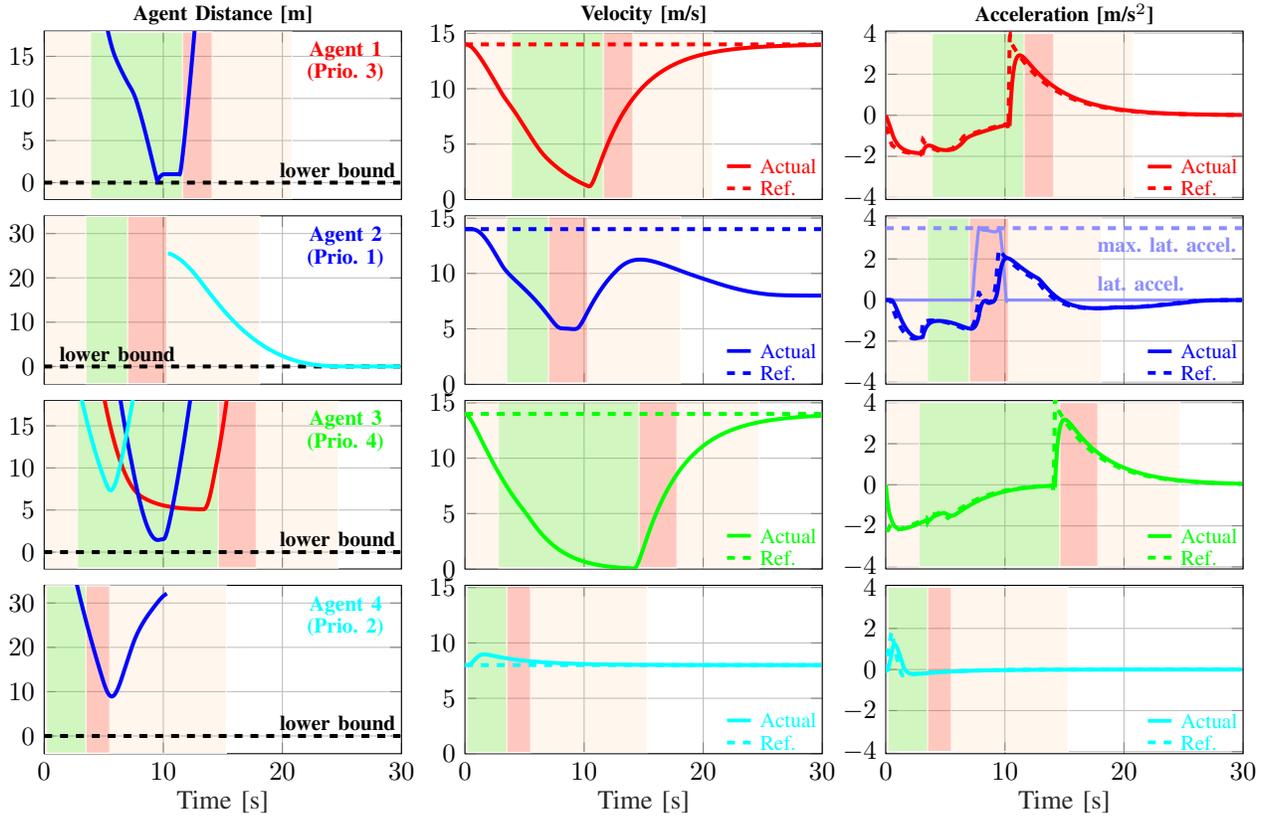}
 		\vspace*{-2mm}
		\caption{From left to right in row $i$: 
		    (1) Distances between the safety region and bounding box of
		    agents $i$ and $l$, 
		    (2) velocity and 
		    (3) acceleration of Agent $i$. 
		    Agent 2 is turning left, therefore, we additionally depict its 
		    lateral acceleration. 
		    Color patches refer to intersection regions: 
		    inside ICR (beige), 
		    BSR (green), 
		    CR (red) 
		    and outside ICR (white).} \vspace*{-9mm}
		\label{fig:results_discussion_resultsPlot}
	\end{center}
\end{figure*}

%---------------------------------------------------------------------------
%---------------------------------------------------------------------------
\subsection{Simulation Setup}
\label{sec:results_setup}
To demonstrate the efficiency of our approach, we present a realistic 
intersection crossing scenario which involves four agents as illustrated in 
\prettyref{fig:results_discussion_snapshots}. 
According to \prettyref{sec:problem_CA}, crossing priorities 
are time-invariant with $\gamma(1)=3$, \mbox{$\gamma(2)=1$}, $\gamma(3)=4$ and 
$\gamma(4)=2$.
In the considered scenario, Agent 1 (red) crosses the intersection straight 
from North to South, Agent 2 (blue) approaches the intersection from the West 
and turns left while Agent 3 (green) and Agent 4 (cyan) cross the intersection 
straight from East to West and South to North, respectively --- as shown in 
\prettyref{fig:results_discussion_snapshots}a.
%Particularly, Agent 1 (red) inhibits the third highest priority and passes the intersection straight from North to South. Agent 2 (blue) has the highest priority and approaches the intersection from the West, turns left and is finally heading towards the North. With the lowest priority, Agent 3 (green) crosses the intersection from East to West. Finally, Agent 4 (cyan) drives from South to North and owns the second highest priority. 

%For every agent, the lenght, width and dynamic drivertrain time constant has been set to $L^{[i]}=\unit[5]{m}$, $W^{[i]}=\unit[2]{m}$ and $T_{a_x}^{[i]}=\unit[0.3]{s}$ while agent priorities are defined as: $\gamma(1)=3$, $\gamma(2)=1$, $\gamma(3)=4$, $\gamma(4)=2$.
All agents have dimensions $L^{[i]}=\unit[5]{m}$ and $W^{[i]}=\unit[2]{m}$ 
and the same dynamic time constant of $T_{a_x}^{[i]}=\unit[0.3]{s}$. 
The initial and reference velocity is $\unitfrac[14]{m}{s}$ for agents 1 to 3 and 
$\unitfrac[8]{m}{s}$ for Agent 4. The initial positions in the global 
frame are: $(-2,82)$ for Agent 1, $(-82,-2)$  for Agent 2, $(69,2)$ for Agent 3 and 
$(2,-39)$ for Agent 4. 
The agents' MPC controllers have the same parameterization: 
a sampling time of $T_s=\unit[0.1]{s}$, a horizon length of $N=50$ and the  
weights $q^{[i]} = q_N^{[i]} = 1$ and $r^{[i]} = 10$. The sampling time is chosen 
in accordance to the commonly applied V2V communication frequency of $\unit[10]{Hz}$. 
We parameterize the safety region by 
\(
      \tilde{d}_{\mathrm{safe,xf/xr}}^{[i]}
{}={}
      \unit[2]{m}
\), 
\(
      \tilde{d}_{\mathrm{safe,yl/yr}}^{[i]}
{}={}
      \unit[1]{m}
\) 
and 
\(
      t_{\mathrm{gap},x/y}^{[i]}
{}={}
      \unit[1]{s}
\). 
Moreover, the longitudinal acceleration is constrained between $-7$ and 
$\unitfrac[4]{m}{s^2}$, the absolute lateral acceleration should not exceed 
$\unitfrac[3.5]{m}{s^2}$ and the total acceleration is bounded 
from above by $\unitfrac[7]{m}{s^2}$. 
The maximum velocity is set to $\unitfrac[15]{m}{s}$. \alex{For equality and inequality constraints, a tolerance of $\epsilon_s=10^{-4}$ has been selected.}
Simulations are run on an Intel i7 machine at $\unit[2.9]{GHz}$ with Matlab R2018b,
while the controllers run in \texttt{C89} using the open source
code generation tool \texttt{nmpc-codegen}, available at \texttt{github.com/kul-forbes/nmpc-codegen}.
%---------------------------------------------------------------------------
%---------------------------------------------------------------------------
%\vspace*{-0.7mm}
\subsection{Discussion of Results}
\label{sec:results_discussion}

\begin{figure*}[t]
	\begin{center}
		\setlength\fwidth{0.80\textwidth}
 		\input{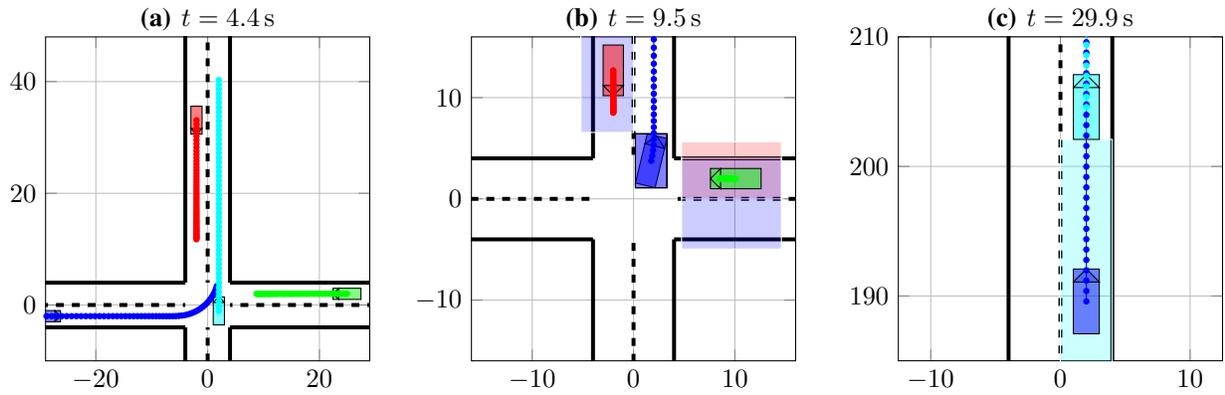}
		\vspace*{-15mm}
		\caption{Snapshots: 
		  (Left) Agent 4 (cyan) crosses first; Note that agents 2 (blue) and 3 (green)  decide to wait;
		  (Middle) Agent 1 (red) and Agent 3 (green) yield to \mbox{Agent 2 (blue);} 
		  (Right) Agent 2 (blue) avoids rear-end collisions with Agent 4 (cyan). 
		  The middle and right figures show the safety region of each agent 
		  $i$ in the color of the conflicting \mbox{Agent $l$.}} 
		\vspace*{-9mm}
		\label{fig:results_discussion_snapshots}
	\end{center}
\end{figure*}

For the given scenario in row $i$, \prettyref{fig:results_discussion_resultsPlot} illustrates  
the resulting trajectories and \prettyref{fig:results_discussion_snapshots} shows three snapshots 
of the scenario along with the agents' optimized trajectories and their safety regions.
As we may observe in \prettyref{fig:results_discussion_resultsPlot} and 
\prettyref{fig:results_discussion_snapshots}, Agent 4 is able to cross the intersection 
before \mbox{Agent 2} by increasing its velocity temporarily from $\unitfrac[8]{m}{s}$ to $\unitfrac[9]{m}{s}$. 
Indeed, the lower priority of Agent 4 allows the agent to cross the intersection 
before Agent 2. After Agent 2 has left the CR, the CA constraint on Agent 2 is dropped by Agent 4 and the distance in the first column is no longer determined. %After leaving the ICR, the CA constraint on Agent 2 is dropped and the distance in the first column is no longer determined. 

While Agent 2 crosses the intersection, it has to decelerate in order to satisfy the 
constraint on its lateral acceleration, which is depicted in the third column (light blue solid) 
along with its upper bound (light blue dashed). 
When passing the CR, no CA constraints are imposed on other agents as \mbox{Agent 2} has the 
highest priority and $\mathcal{A}_{c,\text{ahead}}^{[i]}=\emptyset$. However, after leaving the CR, Agent 2 needs to avoid rear-end collisions 
with Agent 4 who is now driving in the same lane. 
It can be seen that Agent 2 is closing up to Agent 4, however, eventually, Agent 2 is driving 
at the same speed as Agent 4 to avoid a rear-end collision as shown in 
\prettyref{fig:results_discussion_snapshots}c. 

Agent 1, who has lower priority than Agent 2, must  wait for Agent 2 to leave the intersection. Although Agent 1 needs to decelerate to $\unitfrac[1.2]{m}{s}$, it stays as close as possible to \mbox{Agent 2} to maximize its own speed as shown in \prettyref{fig:results_discussion_snapshots}b. 

Lastly, Agent 3 is able to cross. With the lowest priority, it has to impose CA constraints on all other agents. Note that Agent 3 has to wait at the stopping line as it does not %is not able to 
satisfy the minimum spatial preview constraint \prettyref{eq:problem_minPreview_stopCondCons} --- see \prettyref{fig:results_discussion_snapshots}b.
%\begin{figure}[h!]
%	\centering	
%	\hspace*{-9mm}
%	\setlength\fwidth{0.44\textwidth}
%	\input{Figures/scenario_v1.tex}
%	\caption{Intersection scenario. \todo{Snapshot (1) before entering intersection, (2) while passing and (3) read-end collision avoidance}}
%	%	\vspace*{-4mm}
%	\label{fig:results_discussion_snapshots}
%	%	\vspace*{-9mm}
%\end{figure}
After Agent 1 has cleared the intersection, Agent 3 is finally able to proceed. In essence, the distributed intersection coordination scheme satisfies all our 
requirements by being capable of avoiding collisions in a complex realistic driving 
scenario and by ensuring smooth velocity and acceleration trajectories. 
In its current formulation, the control scheme avoids collisions with crossing agents 
in intersections and acts as a conventional adaptive cruise control system (ACC) 
outside intersection areas. Thus, we have proposed a generalized formulation of 
the collision avoidance problem which will serve as a basis for future extensions.

\begin{figure}[ht]
	\vspace*{-4mm}
	\centering	
	%	\hspace*{-5mm}
	\setlength\fwidth{0.34\textwidth}
 	\input{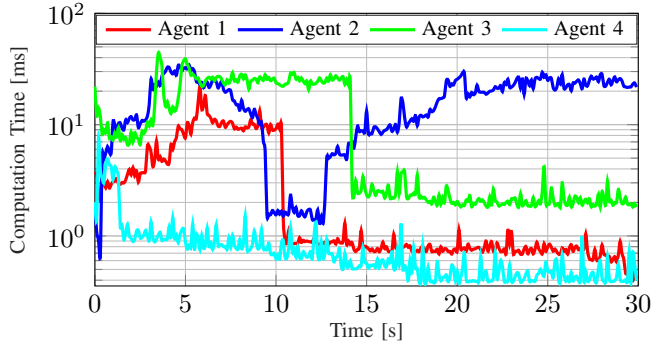} \vspace*{-5mm}	
	\caption{Computation times for every agent for the entire simulation run.} %\alex{The maximum computation time is $\unit[45.1]{ms}$ for Agent 3.}}	
	\label{fig:results_discussion_compTimes}
	\vspace*{-5mm}	
\end{figure}

To evaluate the associated computational complexity,  in \prettyref{fig:results_discussion_compTimes} 
we show the computation times for every agent. Evidently, the entire scheme is real time capable 
having a maximum computation time of \alex{$\unit[44.1]{ms}$}, while the sampling time is 
$\unit[100]{ms}$. An increased computational demand can be recognized when constraints are active for almost the entire horizon. This can be observed for Agent 2 during rear-end collision avoidance ($t \geq \unit[20]{s}$) and for Agent 3 before entering the intersection ($\unit[5]{s} \leq t \leq \unit[14]{s}$).
%For Agent 3, a higher computational demand for rear-end collision avoidance ( $\unit[20]{s} \leq t \leq \unit[30]{s}$).  
%
%For Agent 2, it can be recognized that rear-end collision avoidance is computationally demanding 
%
%
%\markred{Obviously, rear-end collision avoidance is computationally more demanding 
%as it can be recognized for Agent 2. Actually, CA constraints with crossing agents might 
%only be active for a few time instants while rear-end CA constraints are mostly active 
%for the entire horizon.}

%---------------------------------------------------------------------------
%---------------------------------------------------------------------------
%---------------------------------------------------------------------------
% CONCLUSIONS
%---------------------------------------------------------------------------
%---------------------------------------------------------------------------
%---------------------------------------------------------------------------
\section{Conclusion and Future Work}
\label{sec:conclusion}
For the coordination of agents, we have proposed a distributed MPC scheme which relies 
on V2V communication. Every agent determines \alex{its} local control actions by efficiently 
solving a nonconvex nonlinear MPC problem using PANOC. 
To obtain a more general problem formulation, CA constraints are stated in a global 
Cartesian frame using B-spline functions. 
The OCP is augmented with conditional constraints which allows us to encode rules. 
Simulation results demonstrate the efficiency in terms of control performance and 
computational complexity. 

In the future, we envision to further extend the control scheme to multi-lane/multi-intersection 
scenarios and to investigate how to deal with dynamic priority negotiation. 
Finally, we intend to conduct experimental tests. %plan to validate the approach in experimental tests.

%---------------------------------------------------------------------------
%---------------------------------------------------------------------------
%---------------------------------------------------------------------------
% BIBLIOGRAPHY
%---------------------------------------------------------------------------
%---------------------------------------------------------------------------
%---------------------------------------------------------------------------
\vspace*{-1mm}
\bibliographystyle{IEEEtran}
\bibliography{IEEEabrv}

%%%%%%%%%%%%%%%%%%%%%%%%%%%%%%%%%%%%%%%%%%%%%%%%%%%%%%%%%%%%%%%%%%%%%%%%%%%%%%%%
%\section{ACKNOWLEDGMENTS}
%
%The authors gratefully acknowledge the contribution of National Research Organization and reviewers' comments.

%%%%%%%%%%%%%%%%%%%%%%%%%%%%%%%%%%%%%%%%%%%%%%%%%%%%%%%%%%%%%%%%%%%%%%%%%%%%%%%%

\end{document}